\documentclass[11pt,a4paper]{article}

\usepackage{jheppub}
\usepackage{graphicx}
\usepackage{dcolumn}
\usepackage{bm}
\usepackage{amsmath}
\usepackage{braket}
\usepackage{epstopdf}
\usepackage{placeins}
\usepackage{multirow}
\usepackage{makecell}
\usepackage{cleveref}
\usepackage{slashed}

\newcommand{\beq}{\begin{eqnarray}}
\newcommand{\eeq}{\end{eqnarray}}
\newcommand{\beqnn}{\begin{eqnarray*}}
\newcommand{\eeqnn}{\end{eqnarray*}}

\newcommand{\Tr}{\ensuremath{\mathrm{Tr}}}
\newcommand{\SU}{\ensuremath{\mathrm{SU}}}

\newcommand{\MS}{\ensuremath{\overline{\mathrm{MS}}}}
\newcommand{\mutwoGeV}{\mu = 2~\ensuremath{\mathrm{GeV}}}
\renewcommand{\P}{\ensuremath{\mathrm{P}}}
\renewcommand{\S}{\ensuremath{\mathrm{S}}}
\newcommand{\A}{\ensuremath{\mathrm{A}}}
\newcommand{\R}{\ensuremath{\mathrm{R}}}
\newcommand{\W}{\ensuremath{\mathrm{W}}}
\newcommand{\PCAC}{\ensuremath{\mathrm{PCAC}}}
\newcommand{\TEK}{\ensuremath{\mathrm{TEK}}}

\renewcommand\thesection{\arabic{section}}

\usepackage[font=small, labelfont=bf, textfont={small,it}]{caption}
\def\spose#1{\hbox to 0pt{#1\hss}}
\def\ltapprox{\mathrel{\spose{\lower 3pt\hbox{$\mathchar"218$}}
\raise 2.0pt\hbox{$\mathchar"13C$}}}

\begin{document}

\title{The large-$N$ limit of the chiral condensate from twisted reduced models}

\author[a]{Claudio Bonanno,}
\author[a,b]{Pietro Butti,}
\author[a]{Margarita Garc\'ia Per\'ez,}
\author[a,c]{Antonio Gonz\'alez-Arroyo,}
\author[d,e]{Ken-Ichi Ishikawa,}
\author[e]{Masanori Okawa}

\affiliation[a]{Instituto de F\'isica T\'eorica UAM-CSIC, Calle Nicol\'as Cabrera 13-15,\\Universidad Aut\'onoma de Madrid, Cantoblanco, E-28049 Madrid, Spain}

\affiliation[b]{Departamente de F\'isica T\'eorica, Facultad de Ciencias and \\ Centro de Astropart\'iculas y F\'isica de Altas Energ\'ias (CAPA),  
\\ Universidad de Zaragoza, Calle Pedro Cerbuna 12, E-50009, Zaragoza, Spain}

\affiliation[c]{Departamento de F\'isica Te\'orica, Universidad Aut\'onoma de Madrid,\\M\'odulo 15, Cantoblanco, E-28049 Madrid, Spain}

\affiliation[d]{Core of Research for the Energetic Universe,\\Graduate School of Advanced Science and Engineering,\\Hiroshima University, Higashi-Hiroshima, Hiroshima 739-8526, Japan}

\affiliation[e]{Graduate School of Advanced Science and Engineering, Hiroshima University,\\Higashi-Hiroshima, Hiroshima 739-8526, Japan}

\emailAdd{claudio.bonanno@csic.es}
\emailAdd{pbutti@unizar.es}
\emailAdd{margarita.garcia@csic.es}
\emailAdd{antonio.gonzalez-arroyo@uam.es}
\emailAdd{ishikawa@theo.phys.sci.hiroshima-u.ac.jp}
\emailAdd{okawa@hiroshima-u.ac.jp}

\abstract{ We compute the large-$N$ limit of the QCD chiral condensate on the lattice using twisted reduced models, and performing controlled continuum and chiral extrapolations. We perform two different calculations: one consists in extracting the chiral condensate from the quark mass dependence of the pion mass, and the other consists in extracting the chiral condensate from the behaviour of the mode number of the Wilson-Dirac operator for small eigenvalues. We find consistency between the results of the two methods, giving a joint estimate of $\lim_{N\to\infty}\Sigma(N)/N=[184(13)$ MeV$]^3$ ($\overline{\mathrm{MS}}$, $\mu=2$ GeV, taking the square root of the string tension $\sqrt{\sigma}=440$ MeV to set the scale), in remarkable agreement with the $\mathrm{SU}(3)$ 2-flavor FLAG result.}



\maketitle

\flushbottom

\section{Introduction}\label{sec:intro}
In the last few years, there has been intense activity on the study of lattice gauge theories in the limit of large number of colours. Some authors 
address this study by extrapolating the results obtained from small (less than 10) values of $N$~\cite{DelDebbio:2001sj,Lucini:2001ej,DelDebbio:2002xa,Lucini:2004my,DelDebbio:2006yuf,Vicari:2008jw,Allton:2008ty,Lucini:2010nv,Lucini:2012gg,Bali:2013kia,Bonati:2016tvi,Ce:2016awn,DeGrand:2016pur,Hernandez:2019qed,Bennett:2020hqd,DeGrand:2020utq,Hernandez:2020tbc,Bonanno:2020hht,DeGrand:2021zjw,Athenodorou:2021qvs,Bennett:2022gdz,Bonanno:2022yjr,DeGrand:2023hzz}. Our approach, instead, makes use of reduced models~\cite{PhysRevLett.48.1063,BHANOT198247,Gross:1982at,GONZALEZARROYO1983174,PhysRevD.27.2397,Aldazabal:1983ec,Kiskis:2002gr,Narayanan:2003fc,Kovtun:2007py,Unsal:2008ch,Gonzalez-Arroyo:2010omx,Neuberger:2020wpx}, in which the values of $N$ are much larger (over a 100) but the lattice volume is very small and mostly reduces to a single point~\cite{Gonzalez-Arroyo:1983cyv,Das:1984jh,Das:1985xc,Narayanan:2004cp,Gonzalez-Arroyo:2005dgf,Kiskis:2009rf,Hietanen:2009ex,Hietanen:2010fx,Bringoltz:2011by,Hietanen:2012ma,Gonzalez-Arroyo:2012euf,Gonzalez-Arroyo:2013bta,Lohmayer:2013spa,Gonzalez-Arroyo:2014dua,GarciaPerez:2014azn,GarciaPerez:2015rda,Perez:2015ssa,Gonzalez-Arroyo:2015bya,Perez:2017jyq,GarciaPerez:2020gnf,Perez:2020vbn,Butti:2022sgy}. Both methods are indeed complementary. Actually, the finite $N$ corrections of the reduced model approach are of a different nature and one needs to combine with the finite $N$ results of the standard method to extract the genuine $1/N$ corrections to physical observables. In addition, there are observables which are easier to compute in one of the two approaches. For those in which both approaches are available like the string tension or the fundamental meson spectrum of pure Yang-Mills theory, consistent results have been obtained within the quoted precision. This paper can be exactly placed within this context, as it concerns the study of a new observable using the reduced model which has also been extensively studied for $\SU(3)$ using more traditional methods: the chiral condensate. 

The large-$N$ limit exhibits several properties that make it simpler to analyze this quantity compared to finite-$N$ gauge theories. For example, since fermion loops with a finite number of flavours are subleading, the quenched approximation becomes exact at $N=\infty$. In relation to this, also chiral logs are suppressed, and the chiral behaviour of observables is simplified. Moreover, meson resonances become stable and the study of the spectrum is simplified. The reduced model approach encompasses all these properties and, given the large values of $N$ used, corrections to these phenomena are completely negligible. On the negative side, the finite $N$ values largely arise as finite volume effects. For a single-site model, the effective lattice volume equals $N^2$. This forces $N$ to be large enough as the bare coupling $b=1/(g^2 N)$ grows, so as to keep the physical volume $N^2a(b)^4$ large enough. This, together with computational limitations, has dictated our choices for the number of colours. 

In this work we use a subset of configurations that were employed for the study of the fundamental meson spectrum at large $N$~\cite{Perez:2020vbn}. This corresponds to a single site reduced model with twisted boundary conditions: the Twisted Eguchi--Kawai (TEK) model~\cite{GONZALEZARROYO1983174,PhysRevD.27.2397,Gonzalez-Arroyo:2010omx}. The values of $N=289$ and $N=361$ used in this paper correspond effectively to hypercubic lattices of size $17^4$ and $19^4$ respectively. We have employed the standard over-relaxation algorithm~\cite{Perez:2015ssa} to generate configurations, having relatively small auto-correlation times. We used a standard Wilson action with bare coupling $1/b$ in a range that shows a good scaling behaviour of physical quantities~\cite{Perez:2020vbn,Butti:2023hfp,Butti:WiP}. For additional details of the simulations, we refer to Refs.~\cite{Perez:2015ssa,Perez:2020vbn}. 

As mentioned earlier, the present paper focuses on the determination of the fundamental quark condensate for the large-$N$ gauge theory. Several methods have been employed in the literature to obtain the quark condensate for various gauge theories. Most of the determinations available in the literature deal with the case of QCD with $N=3$ colours and various fermion contents~\cite{Giusti:2008vb,JLQCD:2008zxm,Borsanyi:2012zv,Brandt:2013dua,Engel:2014eea,Boyle:2015exm,Wang:2016lsv,Alexandrou:2017bzk,Aoki:2017paw,ExtendedTwistedMass:2021gbo,Liang:2021pql,Bonanno:2023xkg}. On the other hand, the large-$N$ limit has been much less investigated in the literature. In two cases, predictions for the large-$N$ gauge theory have been given in the literature too, although limited to just one lattice spacing~\cite{Narayanan:2004cp,Hernandez:2019qed}. A first exploratory study with continuum-extrapolated results from $N\le 5$ determinations can instead be found in Ref.~\cite{DeGrand:2023hzz}. In this paper, we will provide a novel, solid large-$N$ calculation of the chiral condensate from the TEK model. To ensure the robustness of our determination, we employ 4 different values of the lattice spacing, several values of the valence quark mass, and two rather complementary methods to obtain the condensate, allowing us to perform reliable chiral and continuum extrapolations. Some of the results discussed in this manuscript were also presented at the 2023 Lattice conference~\cite{Bonanno:Pos}.

This paper is organized as follows. In Sec.~\ref{sec:summary_numerical_setup}, after a brief reminder of the TEK model and the parameters entering into its formulation, we describe the two strategies we pursued to compute the condensate. Then, Sec.~\ref{sec:results} contains the results obtained both at finite lattice spacing and in the continuum limit. Some technical aspects are moved to Apps.~\ref{app:ZP} and~\ref{app:additional_plots} to facilitate reading. Finally, in the last Sec.~\ref{sec:conclu} we will present our conclusions.

\section{Numerical setup}\label{sec:summary_numerical_setup}

In this section we will briefly summarize our lattice setup and, most importantly, we will describe the numerical strategies adopted to compute the chiral condensate.

\subsection{Lattice discretization}\label{sec:setup}

All our results have been obtained from the TEK model, i.e., a model involving just $d=4$ $\SU(N)$ matrices with no space-time index. Such a model can be thought of as the reduction on a single-site lattice of an ordinary $d$-dimensional pure Yang--Mills $\SU(N)$ gauge theory defined on a lattice with twisted boundary conditions. Despite its apparent simplicity, there is now plenty of theoretical and numerical evidence that this matrix model is able to correctly reproduce the infinite-volume and infinite-$N$ physics of ordinary Yang--Mills theories~\cite{PhysRevD.27.2397,Gonzalez-Arroyo:2010omx,Gonzalez-Arroyo:1983cyv,Das:1984jh,Das:1985xc,Gonzalez-Arroyo:2012euf,Gonzalez-Arroyo:2014dua,GarciaPerez:2014azn,Gonzalez-Arroyo:2015bya,Perez:2017jyq,GarciaPerez:2020gnf,Perez:2020vbn}. We briefly review the most important details of our model below.

Since in this study we are interested in the large-$N$ limit of QCD with a fixed number of fundamental fermions ('t Hooft limit), i.e., with $N_f/N \to 0$, we only consider a pure-gauge theory in our sea sector, as the contribution of fermions is suppressed as $1/N$ compared to that of the gluons, thus vanishing at $N=\infty$. Therefore, the partition function of the TEK model is just given by:
\beq
Z_\TEK \equiv \prod_{\mu=0}^{d-1} \int [d U_\mu] e^{-S_\TEK[U]},
\eeq
where $[d U_\mu]$ is the invariant $\SU(N)$ Haar measure and
\beq
S_\TEK[U] = -N b \sum_{\nu \ne \mu} z_{\nu\mu} \Tr\left\{U_\mu U_\nu U_\mu^\dagger U_\nu^\dagger\right\}
\eeq
is the TEK Wilson lattice action. Here $b \equiv (g^2 N)^{-1}$ denotes the lattice inverse bare 't Hooft coupling, while $z_{\nu\mu} = z_{\mu\nu}^* = \exp\left\{2\pi i n_{\nu\mu}/N\right\}$ is the twist factor, which is chosen to be an $N^{\text{th}}$ root of $1$.

The choice of the integer-valued anti-symmetric twist tensor $n_{\nu \mu}$ appearing in the twist factor is a crucial aspect in several respects. It should allow for zero-action solutions (orthogonal twist) and avoid zero-modes (irreducible twist). One possible choice is the  so-called symmetric 
twist which demands that we choose  $N=L^2$ to be a perfect square, 
setting  $n_{\nu \mu} = - n_{\mu \nu} = k(L) L$ ($\nu>\mu$), where $k(L)$ is an integer number co-prime with $L$. To avoid center-symmetry breaking~\cite{Ishikawa:2003,Bietenholz:2006cz,Teper:2006sp,Azeyanagi:2007su} one should scale $k(L)$ with $L$ as explained in Refs.~\cite{Gonzalez-Arroyo:2010omx,Chamizo:2016msz,GarciaPerez:2018fkj}. Furthermore, finite $N$ corrections arising from non-planar graphs are reduced with appropriate choices of $k(L)$~\cite{Perez:2017jyq,Bribian:2019ybc}.

Although the fermions in our sea theory are quenched, in this work we will consider a single valence quark to compute the chiral condensate. We discretize the Dirac operator using standard Wilson fermions. After some algebra, the Dirac--Wilson operator in our setup reads~\cite{Gonzalez-Arroyo:2015bya}:
\beq
D_\W^{(\TEK)} = \frac{1}{2\kappa} - \frac{1}{2} \sum_{\mu=0}^{d-1}\left[(\mathbb{I}+\gamma_\mu) \otimes \mathcal{W}_\mu + (\mathbb{I}-\gamma_\mu)\otimes \mathcal{W}_\mu^\dagger \right],
\eeq
where $ 1/(2\kappa) - 4 $ is the unsubstracted bare quark mass, and where $\mathcal{W}_\mu$ is an $N^2 \times N^2$ matrix defined as
\beq
\mathcal{W}_\mu = U_\mu \otimes \Gamma_\mu^*.
\eeq
The $\Gamma_\mu$ are the so-called \emph{twist eaters}, which are $\SU(N)$ matrices satisfying the equation
\beq\label{eq:master_eq_twist_eaters}
\Gamma_\mu \Gamma_\nu = z_{\nu\mu}^* \Gamma_\nu \Gamma_\mu,
\eeq
with $z_{\nu\mu}$ the same twist factor introduced in the TEK action. It can be shown~\cite{Gonzalez-Arroyo:1997ugn} that the choice of a symmetric twist makes the solution of Eq.~\eqref{eq:master_eq_twist_eaters} unique up to a similarity transformation (which is irrelevant when considering observables expressed as traces of link matrices) and multiplication by a center element (which is irrelevant since our action is center symmetric). Thus, we are free to choose any solution of Eq.~\eqref{eq:master_eq_twist_eaters} and keep it fixed to avoid any possible ambiguity.

Finally, we recall that in this work we employed a subset of the gauge configurations generated for the study reported in Ref.~\cite{Perez:2020vbn}. For more details on the ensemble generation, we refer the reader to the original reference.

\subsection{The chiral condensate from the pion mass}

The chiral condensate can be computed from the quark mass dependence of the pion mass by exploiting the so-called Gell-Man--Oakes--Renner (GMOR) relation. At Leading Order (LO) of Chiral Perturbation Theory (ChPT), and assuming the presence of a degenerate quark doublet with mass $m$, it reads:
\beq\label{eq:GMOR}
m^2_\pi = \frac{2 \Sigma m }{F_\pi^2} \equiv 2 B m.
\eeq

In Eq.~\eqref{eq:GMOR} two different Low-Energy Constants (LECs) appear. The first one is the chiral condensate $\Sigma$,
\beq
\Sigma \equiv -\lim_{m\to0}\lim_{V\to\infty} \braket{\overline{\psi}\psi},
\eeq
where $\psi$ refers to a single quark flavour. The second one is the pion decay constant $F_\pi$, which is defined from the following pion matrix element:
\beq\label{eq:Fpi_def}
\sqrt{2} F_\pi \equiv \lim_{m\to0}\lim_{V\to\infty} \frac{2m}{m_\pi^2}\bra{0} \overline{\psi} \gamma_5\psi \ket{\pi(\vec{p}=\vec{0})}.
\eeq
Note that some authors adopt the different definition $f_\pi = \sqrt{2} F_\pi$ for the pion decay constant, which only differs from our definition for an overall numerical factor.

On the lattice, the bare quark mass appearing in the GMOR relation can be defined in several ways, all related to the same renormalized mass by different renormalization factors. Here we consider two possible definitions. One, which renormalizes both multiplicatively and additively, relies on the use of the bare $\kappa$ parameter appearing in the lattice Wilson--Dirac operator:
\beq\label{eq:subtracted_mass}
\frac{1}{2\kappa} - \frac{1}{2\kappa_c} = Z_\S m_\R,
\eeq
where $\kappa_c$ parameterizes the additive renormalization to the bare mass. Another possibility is to use a Ward Identity, the so-called Partially-Conserved Axial Current (PCAC) relation, to define a bare quark mass which only renormalizes multiplicatively:
\beq\label{eq:PCAC_mass}
m_\PCAC = \frac{Z_\P}{Z_\A} m_\R.
\eeq

The pion mass and the pion decay constant, the PCAC mass $m_\PCAC$, the critical $\kappa_c$, the axial renormalization constant $Z_\A$ and the ratio $Z_\P/(Z_\S Z_\A)$ were computed within the TEK model in Ref.~\cite{Perez:2020vbn} from appropriate correlators of the lattice quark propagator ${D_\W^{(\TEK)}}^{-1}$. We refer the reader to that work for more technical details on meson masses computation within the TEK model on the lattice. In the following, we will exploit the results published in that paper to compute the chiral condensate from the GMOR relation.

\subsection{The chiral condensate from the mode number}

The Banks--Casher equation relates the chiral condensate $\Sigma$ to the value of the spectral density at the origin after the chiral limit is taken:
\beq\label{eq:banks_casher}
\frac{\Sigma}{\pi} = \lim_{\lambda\to 0}\lim_{m\to0}\lim_{V\to\infty} \rho(\lambda,m),
\eeq
where $i \lambda + m$ stands for a generic eigenvalue of the massive Dirac operator $\slashed{D} + m$.

In~\cite{Giusti:2008vb}, it has been shown by Giusti and L\"uscher that such relation can be exploited on the lattice to numerically compute the chiral condensate by making use of the mode number of the lattice Dirac operator.

As a matter of fact, this quantity has a very simple relation with the spectral density, being just its integral:
\beq\label{eq:modenumber_def}
\braket{\nu(M)} \equiv \braket{ \# \,\, \vert i\lambda+m \vert \le M} = V \int_{-\Lambda}^{\Lambda} \rho(\lambda,m) d\lambda, \qquad \Lambda^2 \equiv M^2 - m^2.
\eeq
Therefore, it contains the same information about the chiral condensate as the spectral density. In particular, defining an ``effective'' (i.e., quark-mass dependent) chiral condensate as
\beq\label{eq:sigma_eff}
\Sigma \equiv \frac{\pi}{2} \frac{\nu(M)}{V \Lambda},
\eeq
it is easy to show that, in the thermodynamic and chiral limit, Eq.~\eqref{eq:sigma_eff} just reduces to Eq.~\eqref{eq:banks_casher} integrated from $-\Lambda$ to $\Lambda$. Following the lines of~\cite{Giusti:2008vb}, and starting from Eq.~\eqref{eq:sigma_eff}, it is more convenient to express the effective condensate~\eqref{eq:sigma_eff} in terms of renormalized quantities as:
\beq\label{eq:sigma_eff_renorm}
\Sigma_\R = \frac{\pi}{2V} \sqrt{1-\left(\frac{m_\R}{\overline{M}_\R}\right)^2} s_\R,
\eeq
where we used that $\braket{\nu}=\braket{\nu_\R}$~\cite{Giusti:2008vb}, and where $s_\R \equiv \dfrac{\partial \braket{\nu_\R(M_\R)}}{\partial M_\R}\bigg\vert_{M_\R=\overline{M}_\R}$ is the slope of the mode number as a function of the renormalized spectral cut-off $M_\R = M / Z_\P$~\cite{Giusti:2008vb} computed in the point $M_\R = \overline{M}_\R$.

As a matter of fact, looking at the mode number sufficiently close to the origin, and working with a sufficiently large lattice volume and a sufficiently light quark mass, from the Banks--Casher relation one expects the lattice spectral density to develop a plateau, which will translate into a linear rise of $\braket{\nu_\R}$ as a function of $M_\R$. Therefore, the method proposed in~\cite{Giusti:2008vb} consists in estimating $s_\R$ from a linear fit of the mode number in a region close to $M_\R \simeq m_\R$, and then use Eq.~\eqref{eq:sigma_eff_renorm} to estimate the effective condensate from the lattice. Within such an approach, the value of $\overline{M}_\R$ is just the middle point of the fit range.

On the lattice, there are a few ways of computing the mode number, one popular example being noisy estimators~\cite{Giusti:2008vb,Cichy:2015jra}. In this work, we instead directly computed the first few hundred eigenvalues of the Hermitian lattice Dirac--Wilson operator $Q_\W^{(\TEK)} = {Q_\W^{(\TEK)}}^\dagger$,
\beq
Q_\W^{(\TEK)} \equiv \gamma_5 D_\W^{(\TEK)}, \qquad Q_\W^{(\TEK)} u_\lambda = \lambda u_\lambda, \qquad \lambda \in \mathbb{R},
\eeq
using the \href{https://github.com/opencollab/arpack-ng}{\texttt{ARPACK}} library, and obtained $\braket{\nu_\R}$ directly from the counting of the modes lying below a varying threshold. Since we need to count modes below the renormalized cut-off $M_\R = M / Z_\P$, we exploited the knowledge, on the same gauge ensemble used to obtain the Dirac spectra, of the PCAC mass $m_\PCAC$ and of the renormalization constant of the axial current $Z_\A$. Indeed, since $Z_\A m_\PCAC = Z_\P m_\R$, and since a bare eigenvalue renormalizes as $M$, i.e., $\lambda_\R = \lambda / Z_\P$, one can simply consider the renormalized ratio
\beq
\frac{\lambda}{Z_\A m_\PCAC} = \frac{\lambda_\R}{m_\R},
\eeq
and just count the modes with $\lambda_\R/m_\R$ below $M_\R / m_\R$.

\section{Results}\label{sec:results}

In the next subsections we will present our results for the chiral limit of the chiral condensate, obtained for 4 values of the bare coupling $b$ and using both the mode number and the quark mass dependence of the pion mass. Finally, we will perform a continuum extrapolation of our results. Since we expect the following large-$N$ scaling for the condensate:
\beq
\Sigma(N) = N \overline{\Sigma},
\eeq
plus higher-order corrections in $1/N$, in what follows we will always report results for the quantity $\Sigma/N$, which is expected to have a finite large-$N$ limit.

\subsection{Results from the mode number}

Let us start by discussing our results for the chiral condensate obtained from the low-lying spectrum of the Hermitian Dirac--Wilson operator $Q_\W^{(\TEK)}$. In what follows, we will always be referring to results obtained for $N=289$, unless differently stated. We calculated the lowest 300 eigenvalues of $Q_\W^{(\TEK)}$ considering, for each value of $b$, 100 well-decorrelated configurations, separated by 8000 over-relaxation steps. The chosen value of the twist tensor introduced in Sec.~\ref{sec:setup} was $n_{\nu \mu} = k \times L = k \times \sqrt{N} = 5 \times 17$ ($\nu > \mu$).

In order to extrapolate our results for the effective condensate at finite valence quark mass towards the chiral limit, for all values of $b$ we considered 3 different values of the Wilson hopping parameter $\kappa$. Such values were chosen so as to correspond, for all lattice spacings, roughly to the same pion masses. The values of $m_\pi$ and of $m_\PCAC$ employed in the following are all obtained from the chiral fits reported in Ref.~\cite{Perez:2020vbn}, while the scale setting was done according to the determinations of the string tension reported in the same paper. Obtained results for the effective chiral condensate~\eqref{eq:sigma_eff_renorm} from the mode number fits are summarized in Tab.~\ref{tab:eff_chir_cond_finite_m}.

\begin{table}[!t]
\begin{center}
\begin{tabular}{ |c|c|c|c|c|c|}
\hline
$b$ & $\kappa$ & $a m_{\PCAC}$ & $m_\pi/\sqrt{\sigma}$ & $a \sqrt{\sigma}$ & $[\Sigma_\R/(N Z_\P)]^{1/3}$~[MeV] \\
\hline
\multirow{3}{*}{0.355} & 0.1600 & 0.04695(42) & 1.536(6) & \multirow{3}{*}{0.2410(31)} & 255(4)  \\
& 0.1610 & 0.03074(32) & 1.250(4) & & 252(4)  \\
& 0.1615 & 0.02231(37) & 1.050(5) & & 248(4)  \\
\hline
\multirow{3}{*}{0.360} & 0.1580 & 0.03698(44) & 1.536(6) & \multirow{3}{*}{0.2058(25)} & 261(5)  \\
& 0.1588 & 0.02350(61) & 1.250(4) & & 253(5)  \\
& 0.1592 & 0.01646(72) & 1.050(5) & & 248(7)  \\
\hline
\multirow{3}{*}{0.365} & 0.1565 & 0.02580(45) & 1.536(6) & \multirow{3}{*}{0.1784(17)} & 254(4)  \\
& 0.1573 & 0.01319(60) & 1.250(4) & & 245(5)  \\
& 0.1576 & 0.00724(69) & 1.050(5) & & 240(9)  \\
\hline
\multirow{3}{*}{0.370} & 0.1550 & 0.02322(49) & 1.536(6) & \multirow{3}{*}{0.1573(18)} & 250(5)  \\
& 0.1556 & 0.01235(62) & 1.250(4) & & 245(7)  \\
& 0.1559 & 0.00667(70) & 1.050(5) & & 243(10) \\
\hline
\end{tabular}
\end{center}
\caption{Determinations of the effective chiral condensate at finite quark mass as a function of the bare coupling $b$ from the mode number of the Hermitian Dirac--Wilson operator $Q_\W^{(\TEK)}= \gamma_5 D_\W^{(\TEK)}$ for $N=289$ colours. The obtained values of $[\Sigma_\R/(N Z_\P)]^{1/3}$, expressed in units of $\sqrt{\sigma}$, were converted to $\mathrm{MeV}$ units using $\sqrt{\sigma}=440~\mathrm{MeV}$. The values of $a \sqrt{\sigma}$, $m_\pi/\sqrt{\sigma}$ and of $a m_\PCAC$ are taken from Ref.~\cite{Perez:2020vbn}.}
\label{tab:eff_chir_cond_finite_m}
\end{table}

In all cases, it is possible to find a region, close to the threshold $M_\R / m_\R \simeq 1$, where the mode number $\braket{\nu_\R}/N$ linearly grows as a function of the cut-off mass. From a linear fit of $\braket{\nu_R}/N$ as a function of $M_\R/m_\R$ it is possible to extract the slope $s_\R m_\R / N$ (defined in the previous section), from which we can compute the RG-invariant quantity $\Sigma_\R m_\R / (N \sigma^2)$ using Eq.~\eqref{eq:sigma_eff_renorm}, provided that we use $V \sigma^2 = a^4 \sigma^2 N^2$ for the volume, i.e., that we consider our results obtained on a lattice with an effective size $\ell \sqrt{\sigma} = a \sqrt{\sigma} \sqrt{N}$.

In order to get rid of the quark mass, we use the determinations from Ref.~\cite{Perez:2020vbn} of $Z_\A$ and $m_\PCAC$ to compute $Z_\P m_\R / \sqrt{\sigma} = Z_\A m_\PCAC / \sqrt{\sigma}$, so that we are able to express the renormalized condensate in terms of the unknown renormalization constant $Z_\P$. For the moment, we will not take care of this, as we will do the chiral limit at fixed $b$, so this constant will be the same for all pion masses used for the chiral extrapolation. We will however come back to this point when we will discuss the continuum limit extrapolation of the condensate later on.
Finally, we express the third root of $\Sigma_\R / (N Z_\P \sigma^{3/2})$ in MeV units by using the conventional value $\sqrt{\sigma} = 440$ MeV.

\begin{figure}[!t]
\centering
\includegraphics[scale=0.42]{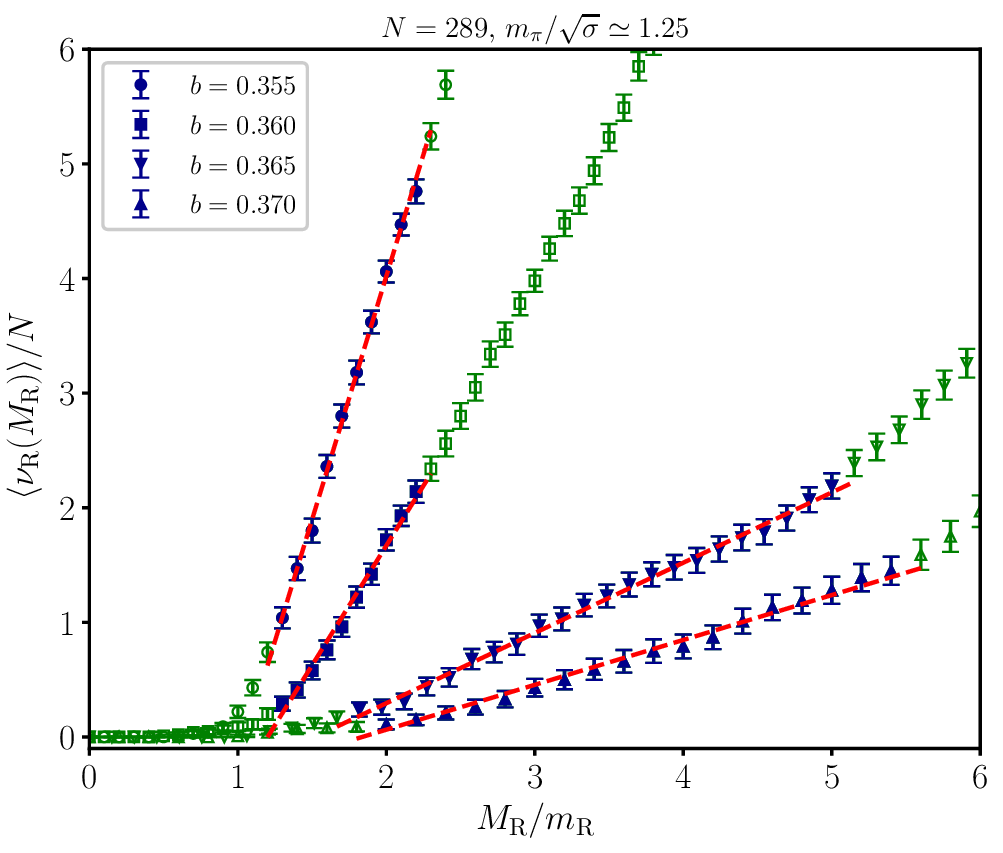}
\includegraphics[scale=0.42]{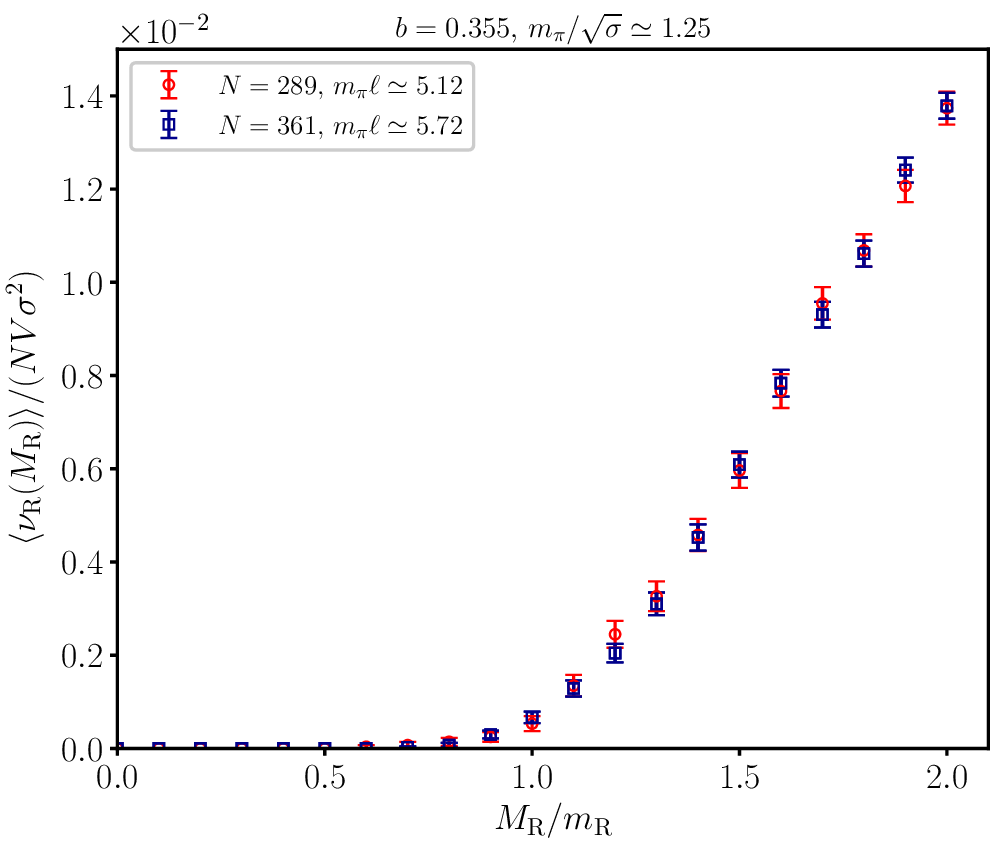}
\caption{Left: behavior of the mode number $\braket{\nu_\R}/N$, computed for $N=289$, as a function of $M_\R/m_\R$ for four values of the bare coupling $b$ and four different values of $\kappa$, approximately corresponding to $m_\pi/\sqrt{\sigma} \simeq 1.25$. Full points correspond to those included in the linear best fit to determine the slope. Right: behavior of the mode number $\braket{\nu_\R}/N$ as a function of $M_\R/m_\R$ obtained using $N=289$ and $N=361$ for the same value of $b=0.355$ and $\kappa=0.1610$. For each value of $N$ we reported the corresponding effective lattice size $\ell = a \sqrt{N}$ in units of the pion mass. In both cases the mode numbers were normalized to the effective volume expressed in units of the string tension: $V\sigma^2 = a^2 N \sigma$.}
\label{fig:linear_fit_nu}
\end{figure}

In the left panel of Fig.~\ref{fig:linear_fit_nu} we show examples of the linear fit to $\braket{\nu_\R(M_\R)}/N$ for 4 different values of the parameters $b$ and $\kappa$. The values of the hopping parameters were chosen so as to correspond, for all lattice spacings, to approximately the same value of the pion mass. In all cases, we observe that a linear function in $M_\R/m_\R$ describes well the behaviour of the mode number close to $M_\R \simeq m_\R$, as expected. Further plots of the mode number are shown in App.~\ref{app:additional_plots}.

In order to verify that our calculations are not affected by significant finite-size effects, for a particular value of $b$ and $\kappa$ we also repeated the calculation of the mode number using spectra obtained for a larger value of $N$, namely $N=361$. In this case, the mode number was obtained computing the first 300 lowest eigenvalues of $Q_\W^{(\TEK)}$ for an ensemble of 100 well-decorrelated gauge configurations, separated by 12000 over-relaxation steps, and generated choosing $n_{\nu \mu} = k \times \sqrt{N} = 7 \times 19$ ($\nu > \mu$) as the twist tensor. Since we found perfectly agreeing results for the mode number, we conclude that finite-size effects are negligible within our level of precision, cf.~the right panel of Fig.~\ref{fig:linear_fit_nu}.

Being the determinations of the effective chiral condensate shown so far obtained at finite valence quark mass, it is necessary to extrapolate them towards the chiral limit. As worked out in Ref.~\cite{Giusti:2008vb} using ChPT, the effective chiral condensate is expected to receive linear corrections in the valence quark mass at the lowest order. Chiral logs appear at higher order in the quark mass, but they are expected to drop out in the large-$N$ limit.

\begin{figure}[!t]
\centering
\includegraphics[scale=0.62]{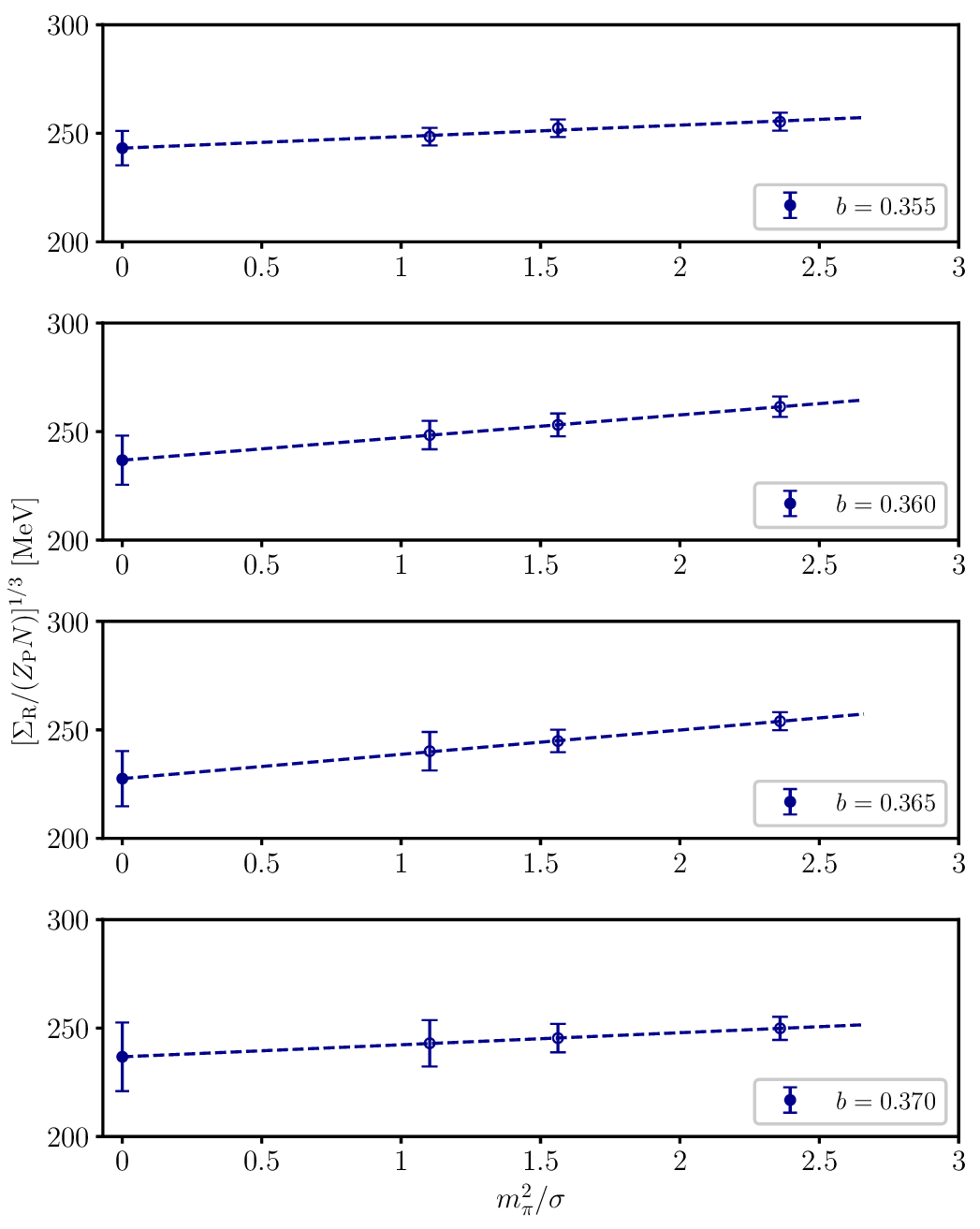}
\caption{Chiral extrapolation at fixed lattice spacing of the effective chiral condensate for $N=289$ for all values of the bare coupling $b$ explored in this work.}
\label{fig:chircond_modenumber_vs_m}
\end{figure}

On the basis of such considerations, and using the GMOR relation, we thus expect the effective chiral condensate to depend linearly on $m_\pi^2$. As a matter of fact, for all explored values of $b$, we find that our data are perfectly described by a law of the type:
\beq
\frac{\Sigma_\R(b,m^2_\pi)}{N Z_\P(b)} = \frac{\Sigma_\R(b)}{N Z_\P(b)} + c(b)\,\frac{m^2_\pi}{\sigma},
\eeq
where $c(b)$ appears to be very mildly dependent on $b$. Chiral extrapolations of the effective condensate are reported in Tab.~\ref{tab:chir_cond_vs_a} and shown in Fig.~\ref{fig:chircond_modenumber_vs_m}.

As a final remark, we would like to stress that we relied on the string tension $\sigma$ for the scale setting, rather than on other observables such as $t_0$, because the error on the scale was sub-dominant in our calculations. In particular, our dominant source of statistical error comes from the PCAC mass and from the slope of the mode number. As an example, for $b=0.370$ and $\kappa=0.1559$, we are able to determine the slope $s_\R m_\R$ with a $\sim 6\%$ precision and $m_\PCAC$ with a $\sim 10\%$ accuracy, while the lattice spacing in units of the string tension is known with a $\sim 1\%$ precision.

\subsection{Results from the pion mass and comparison with the spectral method}

We now move to the determination of the chiral condensate from the GMOR relation. To this end, we can rely on the several determinations of the pion mass as a function of the Wilson hopping parameter $\kappa$ reported in~\cite{Perez:2020vbn}.

More precisely, using the GMOR relation in Eq.~\eqref{eq:GMOR} and the definition of the subtracted quark mass in Eq.~\eqref{eq:subtracted_mass}, it is easy to recognize that the results for the slopes of the squared pion mass $m^2_\pi / \sigma$ as a function of  $1/(2\kappa)$ reported in Tab.~2 of Ref.~\cite{Perez:2020vbn} in units of $\sqrt{\sigma}$ are just $2 \Sigma_\R / (Z_\S \sqrt{\sigma} F_\pi^2)$.

Using the determinations for $Z_\A$, $Z_\P/(Z_\S Z_\A)$ and $F_\pi$ reported in that very same work, it is easy to extract $\Sigma_\R / (N Z_\P \sigma^{3/2})$ from these data, which are then converted in MeV units using $\sqrt{\sigma} = 440$ MeV after taking the third root. These results can then be directly compared, lattice spacing by lattice spacing, with the determinations obtained from the mode number in the previous subsection.

\begin{table}[!t]
\begin{center}
\begin{tabular}{ |c|c|c|c|}
\hline
$b$ & $a \sqrt{\sigma}$ & \makecell[c]{$[\Sigma_\R/(N Z_\P)]^{1/3}$~[MeV]\\(from $\braket{\nu_\R}$)} & \makecell[c]{$[\Sigma_\R/(N Z_\P)]^{1/3}$~[MeV]\\(from $m_\pi$)} \\
\hline
0.355 & 0.2410(31) & 243(8)  & 220(14) \\
0.360 & 0.2058(25) & 237(10) & 220(15) \\
0.365 & 0.1784(17) & 228(13) & 213(14) \\
0.370 & 0.1573(18) & 237(16) & 215(15) \\
\hline
\end{tabular}
\end{center}
\caption{Determinations of the effective chiral condensate extrapolated at zero quark mass as a function of the bare coupling $b$ from the mode number of the Hermitian Dirac--Wilson operator $Q_\W^{(\TEK)}= \gamma_5 D_\W^{(\TEK)}$ and from the quark mass dependence of the pion mass. In both cases, results obtained in units of $\sqrt{\sigma}$ where converted to $\mathrm{MeV}$ units using the conventional value $\sqrt{\sigma}=440$ $\mathrm{MeV}$.}
\label{tab:chir_cond_vs_a}
\end{table}

\begin{figure}[!t]
\centering
\includegraphics[scale=0.62]{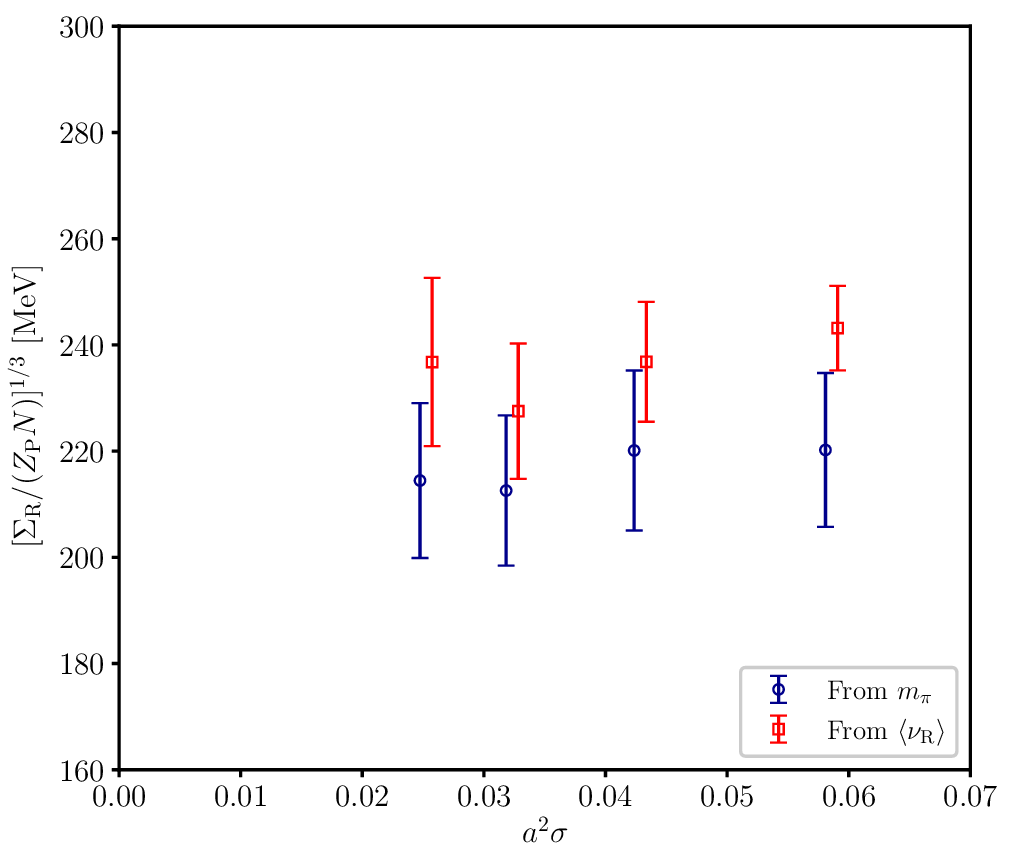}
\caption{Comparison of different determinations of the chiral limit of the effective chiral condensate at four different values of the lattice spacing. Results obtained from the pion mass are plotted slightly shifted to improve the readability of the figure.}
\label{fig:chircond_chiral_limit_comp}
\end{figure}

Note that, while for $Z_\A$ and $Z_\P/(Z_\S Z_\A)$ we simply employed the determinations available for the 4 values of $b$ considered in this work for $N=289$, for the pion decay constant $F_\pi$ we directly used the value:
\beq\label{eq:Fpi_num_res}
\lim_{N\to\infty}\frac{F_\pi}{\sqrt{\sigma}} \sqrt{\frac{3}{N}} = 0.215(21).
\eeq
This value, reported in Tab.~10 in Ref.~\cite{Perez:2020vbn}, is a determination obtained in the large-$N$ and chiral limit from the fit of various determinations of the pion decay constant, obtained for several values of $b$ and $\kappa$, and for two values of $N$, namely $N=289$ and $361$\footnote{Note that the definition of $F_\pi$ used in this work differs from the one adopted in Ref.~\cite{Perez:2020vbn} by a factor of $\sqrt{3/N}$, cf.~Eq.~\eqref{eq:Fpi_def} in this work and Eq.~(4.2) in Ref.~\cite{Perez:2020vbn}.}. On passing, we note that, by using again $\sqrt{\sigma}=440$ MeV, Eq.~\eqref{eq:Fpi_num_res} yields $F_\pi / \sqrt{N} \simeq 55(5)$ MeV, which is in excellent agreement with the large-$N$ determination $F_\pi / \sqrt{N} \simeq 56(5)$ MeV of Ref.~\cite{Bali:2013kia}, obtained with standard methods for a single lattice spacing and extrapolating results obtained for $N\le 17$ towards the large-$N$ limit.

The determinations obtained from the pion mass are reported in Tab.~\ref{tab:chir_cond_vs_a}, along with the chiral limit extrapolations of the mode number ones of the previous subsection. For the sake of comparison, we also plot them together in Fig.~\ref{fig:chircond_chiral_limit_comp}. As it can be appreciated, we find a perfect agreement for all explored lattice spacings, confirming the consistency of the two methods.

\FloatBarrier

\subsection{Continuum limit}

In order to extrapolate our determinations of $[\Sigma_\R/(N Z_\P)]^{1/3}$ towards the continuum limit, it is necessary to get rid of the renormalization constant $Z_\P$. We do not have a determination on our gauge ensemble of such quantity, therefore, we tried to estimate it in two ways. Since $Z_\P$ is scheme and scale-dependent, we need to fix a renormalization scheme and a renormalization point to compute it. In the following discussion, our renormalized results will always refer to the $\MS$ scheme at a renormalization scale of $\mutwoGeV$.

One way of estimating $Z_\P$ consists in using the non-perturbative determinations of $Z_\S$ as a function of the lattice spacing obtained in Ref.~\cite{Castagnini:2015ejr} for $N\le 7$ using the well-established Rome-Southampton method~\cite{Martinelli:1994ty}. Such determinations can be extrapolated towards the large-$N$ limit and then interpolated as a function of $a \sqrt{\sigma}$ to provide results for $Z_\S$ at the lattice spacings employed in this work. Finally, we can convert $Z_\S$ to $Z_\P$ using the non-perturbative determinations obtained from the TEK model of $Z_\P/(Z_\S Z_\A)$ and $Z_\A$ of Ref.~\cite{Perez:2020vbn}.

Another way of estimating $Z_\P$ is instead to use the $1$-loop perturbative results obtained at large-$N$ for Wilson fermions in~\cite{Skouroupathis:2007jd,Skouroupathis:2008mf} and reported in the appendix of Ref.~\cite{Bali:2013kia} (see also the appendix of~\cite{Castagnini:2015ejr}, where the procedure to run $Z_\P$ from the lattice scale $\mu = 1/a$ to the conventional scale $\mutwoGeV$ is explained in detail). However, it is well known numerically that perturbation theory overestimates renormalization constants with respect to the lattice non-perturbative determinations (see, e.g., Ref.~\cite{Perez:2020vbn}, where perturbative and non-perturbative determinations for the constant $Z_\A$ are compared).

Also in our calculations, perturbative determinations of $Z_\P$ turn out to be larger than the non-perturbative ones by about a factor of $\sim 1.6$. We also tried to use several improved bare couplings, instead of just $1/b$, to improve our perturbative estimates of $Z_\P$. Although such determinations all move closer to the non-perturbative ones, as expected, we observe that they still overshoot the non-perturbative results by factors of $\sim 1.3$ -- $1.5$.

In conclusion, to extrapolate our results towards the continuum limit, we rely on the non-perturbative determinations of $Z_\P$ of~\cite{Castagnini:2015ejr} to renormalize the chiral condensate. In any case, we report and compare all our calculations of $Z_\P$ in App.~\ref{app:ZP}. The values of the renormalized condensate are reported in Tab.~\ref{tab:renorm_condensate}, along with the employed values of $Z_\P$.

\begin{table}[!t]
\begin{center}
\begin{tabular}{ |c|c|c|c|c|}
\hline
$b$ & $a\sqrt{\sigma}$ & \makecell{$Z_\P$\\(Non-perturbative)} & \makecell{$(\Sigma_\R/N)^{1/3}$ [MeV]\\(From $\braket{\nu_\R}$)} & \makecell{$(\Sigma_\R/N)^{1/3}$ [MeV]\\(From $m_\pi$)} \\
\hline
0.355 & 0.2410(31) & 0.5212(60) & 196(6)  & 177(11) \\
0.360 & 0.2058(25) & 0.536(12)  & 192(8)  & 178(12) \\
0.365 & 0.1784(17) & 0.555(12)  & 187(10) & 175(12) \\
0.370 & 0.1573(18) & 0.583(20)  & 198(13) & 180(12) \\
\hline
\end{tabular}
\end{center}
\caption{Summary of the renormalized determinations of the chiral condensate employed in our continuum extrapolations. We also report the values of $Z_\P$ used to renormalize our bare results (see the text and App.~\ref{app:ZP} for a detailed discussion on their determination).}
\label{tab:renorm_condensate}
\end{table}

\begin{figure}[!htb]
\centering
\includegraphics[scale=0.41]{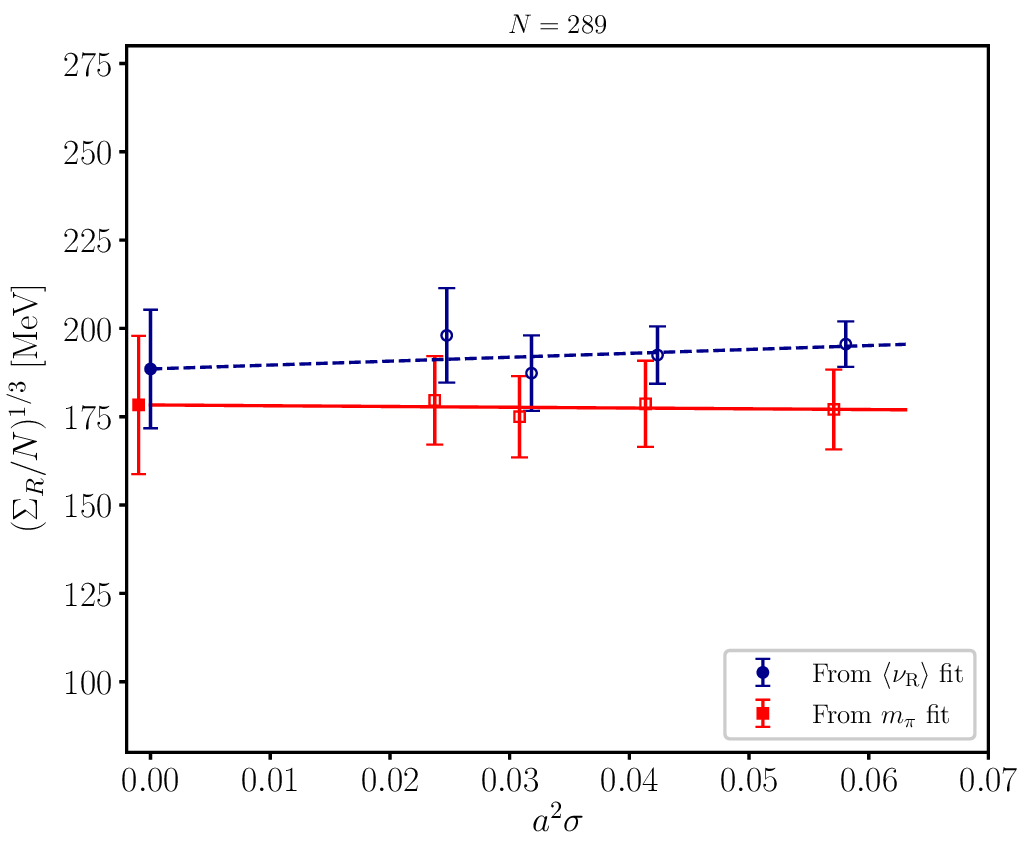}
\includegraphics[scale=0.41]{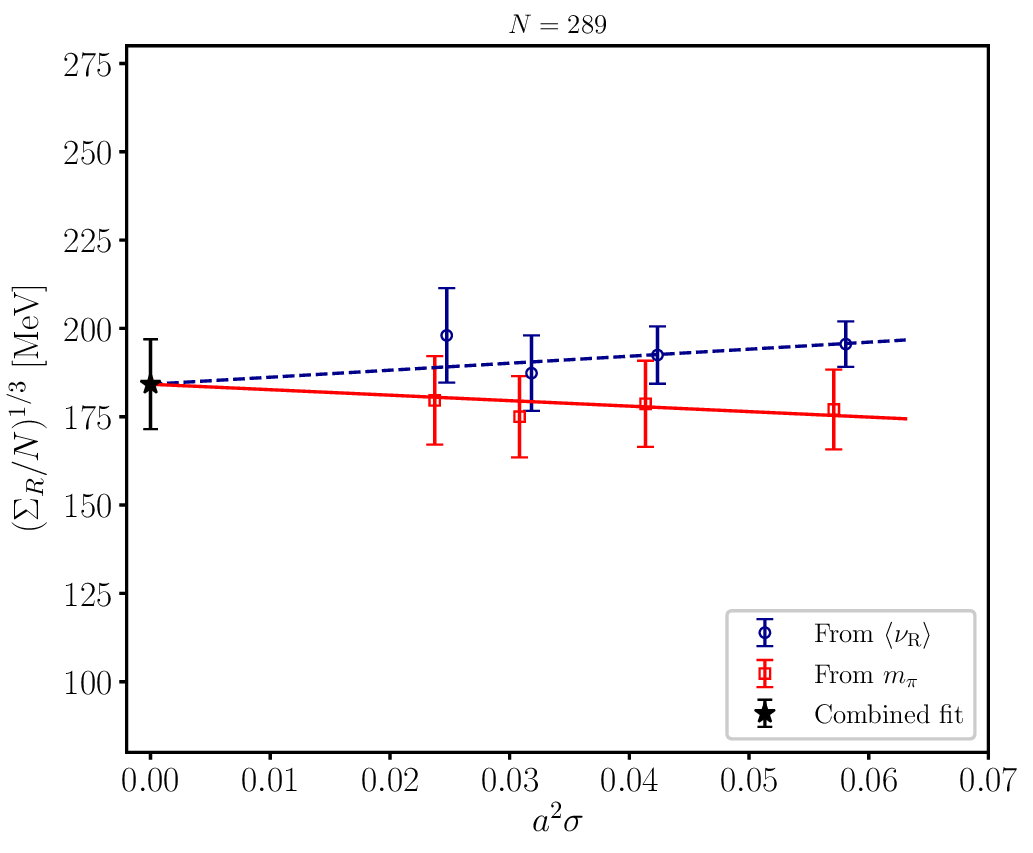}
\caption{Continuum extrapolation of our finite-lattice-spacing determinations of $\Sigma_\R/N$ obtained from the fit of the mode number $\braket{\nu_\R}$ and from the fit of the pion mass $m_\pi$.}
\label{fig:cont_limit}
\end{figure}

As it can be appreciated from Fig.~\ref{fig:cont_limit}, our data can be perfectly fitted to a behaviour of the type:
\beq
\left(\frac{\Sigma_\R}{N}\right)^{1/3}(a\sqrt{\sigma}) = \left(\frac{\Sigma_\R}{N}\right)^{1/3} + k\, a^2 \sigma,
\eeq
and lattice artefacts turn out to be very small with respect to the typical size of our error bars. The best fit of the mode number data yields a continuum limit of
\beq
\left(\frac{\Sigma_\R}{N}\right)^{1/3} = 189(17)~\mathrm{MeV}, \qquad \text{(from $\braket{\nu_\R}$ data),}
\eeq
while the best fit of the pion data yields a perfectly compatible result
\beq
\left(\frac{\Sigma_\R}{N}\right)^{1/3} = 178(20)~\mathrm{MeV} \qquad \text{(from $m_\pi$ data)}.
\eeq
Given the perfect agreement between these two determinations, we also did a combined fit of the two sets of data, obtaining:
\beq\label{eq:FINAL_RESULT}
\left(\frac{\Sigma_\R}{N}\right)^{1/3} = 184(13)~\mathrm{MeV} \qquad \text{(combined fit)},
\eeq
which we quote as our final result for the large-$N$ limit of $(\Sigma_\R/N)^{1/3}$ in the $\MS$ scheme at $\mutwoGeV$.

It is now interesting to compare our result with other determinations in the literature. First of all, we compare with the value obtained for $\SU(3)$ in the 2-flavor theory. The latest FLAG review~\cite{FlavourLatticeAveragingGroupFLAG:2021npn} reports the world average $\Sigma_\R^{1/3} = 266(10)$ MeV for $N_f=2$, which would lead to $(\Sigma_\R/N)^{1/3} \simeq 184(7)$ MeV, which is in remarkable agreement with our final large-$N$ result in Eq.~\eqref{eq:FINAL_RESULT}. This extremely interesting finding points out that higher-order corrections in $1/N$ are pretty small, and cannot be observed at the $4$--$7\%$ level of precision of these determinations. This conclusion fits very well with the fast convergence of the large-$N$ limit observed also for other quantities, such as the coefficients parametrizing the $\theta$-dependence of the vacuum energy~\cite{Bonanno:2020hht}, glueball masses~\cite{Athenodorou:2021qvs} or the critical deconfinement temperature~\cite{Lucini:2005vg}.

Concerning instead other large-$N$ determinations, two results can be found in Refs.~\cite{Narayanan:2004cp,Hernandez:2019qed}. These calculations rely on 1 lattice spacing, and hence no continuum limit can be extracted. Another recent large-$N$ determination can be found in Ref.~\cite{DeGrand:2023hzz}. There, the authors perform a continuum limit using 4 rather coarse lattice spacings, and perform an exploratory study of the large-$N$ limit from results obtained for $N\le 5$. Despite these caveats, we believe that it is still interesting to compare these determinations with our numbers.

In Ref.~\cite{Narayanan:2004cp}, the authors consider an extended hypercubic $L^4$ lattice with $L=6,8$ and periodic boundary conditions and use just one value of $b=0.350$, and a few values of $N$ of the order of a few tens. Their preliminary estimate for the condensate, extracted from the probability distribution of the smallest eigenvalue of the massless overlap operator, and renormalized using a perturbative estimate for $Z_\S$, is $\Sigma_\R/N \approx \left(174~\text{MeV}\right)^3$, which is very close and compatible with our more reliable computation.

In Ref.~\cite{Hernandez:2019qed}, instead, the authors consider a full QCD computation with $N_f=4$ degenerate flavours of Wilson quarks and $N$ varying from 3 to 6. They use a single value of the lattice spacing, $a \simeq 0.075$ fm, and several values of the quark mass. Their estimation of the large-$N$ limit of the bare chiral condensate is $aB_R/Z_\P=1.72(7)$. Using the value $t_0/a^2=3.69(13)$ reported in the paper we get $\sqrt{t_0}B_\R/Z_\P = \sqrt{t_0}\Sigma_\R/(F_\pi^2 Z_\P) \simeq 3.30(15)$. Such number, however, cannot be directly compared with our determination in Tab.~\ref{tab:chir_cond_vs_a} for $b=0.370$ (which corresponds to a similar value of $a$), due to the different regularization employed (the authors of~\cite{Hernandez:2019qed} adopted the Iwasaki gauge action and improved Wilson fermions). Nonetheless, using our chiral limit result for $[\Sigma_\R/(N Z_\P)]^{1/3} = 237(16)$ MeV in Tab.~\ref{tab:eff_chir_cond_finite_m} for $b=0.370$, the result for $F_\pi$ in Eq.~\eqref{eq:Fpi_num_res}, and the conversion factor $\sqrt{8 t_0 \sigma} = 1.078(9)$ computed at large-$N$ in~\cite{Perez:2020vbn}, we obtain the compatible value $\sqrt{t_0} B_R/Z_\P = 3.87(77)$.

Finally, in Ref.~\cite{DeGrand:2023hzz}, the authors perform simulations with $N_f=2$ degenerate flavors of clover-Wilson fermions and Wilson gauge action, considering several values of the quark masses and $N=3,4,5$. The authors also consider 4 values of the bare coupling, and thus are able to provide continuum-extrapolated results. Although the authors of~\cite{DeGrand:2023hzz} acknowledge that the rather small range of $N$-values employed makes it difficult to provide a definitive large-$N$ extrapolated value, using our result for $\Sigma_\R/N$ in Eq.~\eqref{eq:FINAL_RESULT}, the result for $F_\pi$ in Eq.~\eqref{eq:Fpi_num_res}, and the large-$N$ conversion factor $\sqrt{8 t_0 \sigma} = 1.078(9)$~\cite{Perez:2020vbn}, we obtain $\sqrt{t_0} B = \sqrt{t_0} \Sigma_\R/F_\pi^2 = 1.81(36)$, which is compatible with the large-$N$ extrapolated result quoted in the conclusions of~\cite{DeGrand:2023hzz}, $\sqrt{t_0} B = 1.72(22)$.

\FloatBarrier

\section{Conclusions}\label{sec:conclu}

In this work, we have presented a numerical determination of the large-$N$ limit of the QCD chiral condensate from TEK models on the lattice.

Such determination was obtained in the quenched theory with $N=289$ number of colors (although we also checked that results obtained for $N=361$ were fully compatible), and was achieved following two different and complementary strategies, both based on ChPT predictions. One consisted of fitting the quark mass dependence of the pion mass with a linear function, whose slope is related to the ratio of LECs $\Sigma/F_\pi^2$ from the GMOR relation. The other one consisted of using the Giusti--L\"uscher method to extract the chiral condensate from a linear fit of the behaviour of the mode number as a function of the spectral cut-off $M$, whose slope can be related to the chiral condensate from the Banks--Casher relation.

We have shown that, after checking finite volume effects, and after performing controlled chiral and continuum extrapolations using 4 values of the lattice spacing $a$ and several values of the valence quark mass for each $a$, both strategies give perfectly consistent results, leading in particular to our final combined result (obtained assuming $\sqrt{\sigma}=440$ MeV):
\beq
\lim_{N\to\infty} \frac{\Sigma_\R(N)}{N} = \left[184(13)~\text{MeV}\right]^{3}, \qquad (\MS~\text{scheme},\,\, \mutwoGeV).
\eeq
This result is in remarkable agreement with the FLAG world-average for $(\Sigma_\R/N)^{1/3}$ for $N=3$ in the 2-flavor theory, 184(7)~MeV. We also find agreement with the few previous large-$N$ predictions that could be retrieved in the literature.

In the future, we would like to extend our calculation to other interesting physical cases, such as the SUSY case with 1 adjoint Majorana fermion, where the computation of the gluino condensate is of great theoretical interest.

\section*{Acknowledgements}
It is a pleasure to thank Carlos Pena for useful discussions. This work is partially supported by the Spanish Research Agency (Agencia Estatal de Investigaci\'on) through the grant IFT Centro de Excelencia Severo Ochoa CEX2020-001007-S, funded by MCIN/AEI/10.13039 /501100011033, and by grant PID2021-127526NB-I00, funded by MCIN/AEI/10.13039/ 501100011033 and by “ERDF A way of making Europe”. We also acknowledge support from the project H2020-MSCAITN-2018-813942 (EuroPLEx) and the EU Horizon 2020 research and innovation programme, STRONG-2020 project, under grant agreement No 824093. P.~B.~is supported by Grant PGC2022-126078NB-C21 funded by MCIN/AEI/ 10.13039/501100011033 and “ERDF A way of making Europe”. P.~B.~also acknowledges support by Grant DGA-FSE grant 2020-E21-17R Aragon Government and the European Union - NextGenerationEU Recovery and Resilience Program on ‘Astrofísica y Física de Altas Energías’ CEFCA-CAPA-ITAINNOVA. K.-I.~I.~is supported in part by MEXT as "Feasibility studies for the next-generation computing infrastructure". M.~O.~is supported by JSPS
KAKENHI Grant Number 21K03576. Numerical calculations have been performed on the \texttt{Finisterrae~III} cluster at CESGA (Centro de Supercomputaci\'on de Galicia). We have also used computational resources of Oakbridge-CX at the University of Tokyo through the HPCI System Research Project (Project ID: hp230021 and hp220011).

\appendix
\section{Determinations of the renormalization constant $Z_\P$}\label{app:ZP}

In this appendix, we present our calculations to estimate the renormalization constant $Z_\P$, necessary to renormalize the bare chiral condensate. We stress again that we chose the $\MS$ renormalization scheme and fixed the renormalization scale to be $\mutwoGeV$.

First of all, let us present the non-perturbative estimation that we actually used in our work to extrapolate the chiral condensate towards the continuum limit. In Ref.~\cite{Castagnini:2015ejr} several non-perturbative determinations of renormalization constants, obtained using the Rome-Southampton method~\cite{Martinelli:1994ty}, are presented. Their numerical setup consists of standard lattice quenched simulations with Wilson valence quarks. In particular, they report results for the renormalization constant $Z_\S$ for 4 values of the lattice spacing $a\sqrt{\sigma}$, and for a few values of the number of colors $N$, ranging from 3 to 7.

\begin{figure}[!htb]
\centering
\includegraphics[scale=0.44]{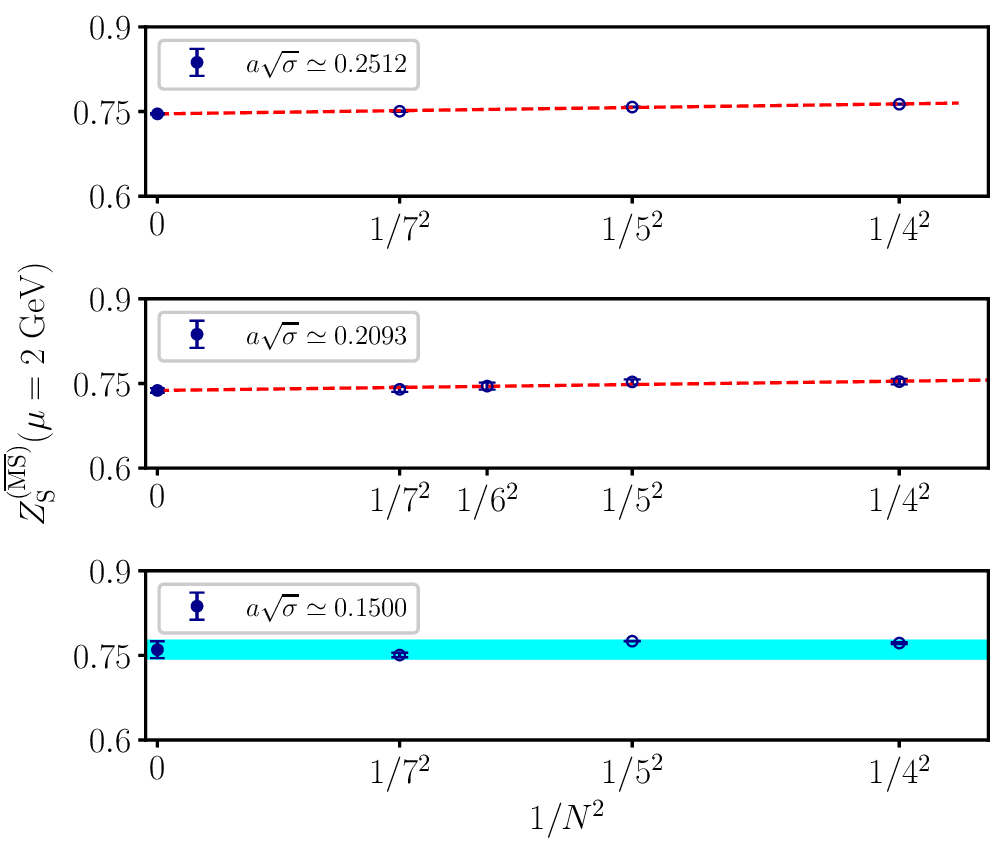}
\includegraphics[scale=0.44]{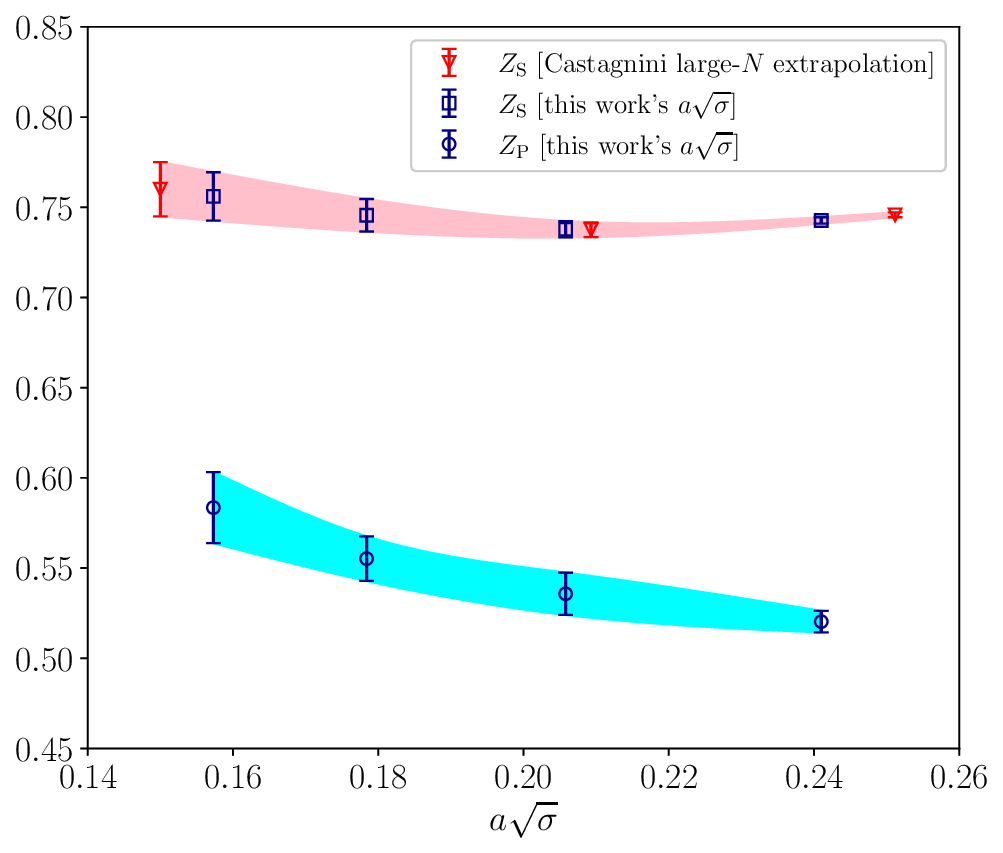}
\caption{Left: large-$N$ extrapolation of the non-perturbative determinations of $Z_\S$ reported in Ref.~\cite{Castagnini:2015ejr} using $N\le 7$ data. Right: interpolation of large-$N$ determinations of $Z_\S$ as a function of $a \sqrt{\sigma}$ at the values of the lattice spacing employed in this work, and related results for $Z_\P$ obtained from the large-$N$ non-perturbative determinations of $Z_\P/Z_\S$ of Ref.~\cite{GarciaPerez:2020gnf} within the TEK model.}
\label{fig:ZP_nonperturbative}
\end{figure}

We extrapolated, using determinations for $N\le 7$, $Z_\S$ towards the large-$N$ limit assuming $1/N^2$ corrections for the 3 lattice spacings relevant for the range explored in this work. For the finest one, since we could not obtain a reliable best fit, we opted for a very conservative estimation of the large-$N$ limit and just took a confidence band including all available determinations in~\cite{Castagnini:2015ejr}. All large-$N$ extrapolations are shown in Fig.~\ref{fig:ZP_nonperturbative} on the left.

Then, we interpolated the obtained large-$N$ estimates of $Z_\S$ as a function of $a\sqrt{\sigma}$ using a cubic spline to obtain them for the lattice spacings considered in this work. Finally, we obtained $Z_\P$ by multiplying such quantities for our non-perturbative determinations of $Z_\P/Z_\S$ obtained from the TEK model~\cite{Perez:2020vbn}. Our final results for $Z_\P$, as well as the interpolation of $Z_\S$, are shown in Fig.~\ref{fig:ZP_nonperturbative} on the right.

Now we move to our calculations of $Z_\P$ from perturbation theory. Let us stress that we did \emph{not} use any of such values anywhere in this paper, and we just show them here for the sake of completeness, and to allow a comparison with the non-perturbative estimations earlier discussed.

Perturbation theory predicts, for unimproved Wilson valence fermions in a quenched sea, and for the $\MS$ scheme at a scale $ \mu = 1/a$, the following perturbative expansion for $Z_\P$ at large-$N$~\cite{Skouroupathis:2007jd,Skouroupathis:2008mf} (cf.~also the appendix of~\cite{Bali:2013kia}):
\beq\label{eq:ZP_pert_2loop}
Z_\P^{(\MS)} ( \mu = 1/a) = 1 &-& 0.580734161(4) \frac{\lambda}{4\pi} - 0.8420(2) \left(\frac{\lambda}{4\pi}\right)^2 + O(\lambda^3),
\eeq
where $\lambda=1/b$ is the lattice bare 't Hooft coupling.

In order to run this quantity from the lattice scale $ \mu_0=1/a$ to the conventional scale $\mutwoGeV$, we use the following relation~\cite{Alexandrou:2012mt}:
\beq
Z_\P^{(\MS)}( \mu) = R( \mu, \mu_0) \, Z_\P^{(\MS)}(\mu_0),
\eeq
where
\beq
R( \mu,\mu_0) = \frac{ \exp\left\{F\left(\frac{\lambda_\R( \mu)}{16\pi^2}\right)\right\} }{ \exp\left\{F\left(\frac{\lambda_\R(\mu_0)}{16\pi^2}\right)\right\} }.
\eeq
The function $F(x)$ and the renormalized running 't Hooft coupling $\lambda_\R(\mu)$ at 3-loop read~\cite{Alexandrou:2012mt}:
\beq
\begin{aligned}
F(x) &= \frac{\gamma_0}{2\beta_0}\log(x) + \frac{\beta_0\gamma_2 - \beta_2\gamma_0}{4\beta_0\beta_2}\log(\beta_0 + \beta_1 x + \beta_2 x^2) \,+\\
\, & \,+\frac{2 \beta_0 \beta_2 \gamma_1 - \beta_1 \beta_2 \gamma_0 - \beta_0 \beta_1 \gamma_2 }{2\beta_0\beta_2\sqrt{4\beta_0\beta_2-\beta_1^2}}\arctan\left(\frac{\beta_1+2 \beta_2 x}{\sqrt{4 \beta_0 \beta_2 -\beta_1^2}}\right);
\end{aligned}
\eeq
\beq
\begin{aligned}
\frac{\lambda_\R(\mu)}{16\pi^2} &= \frac{1}{2 \beta_0 \log(\mu/\Lambda_{\MS})} - \frac{\beta_1}{\beta_0^3}\frac{\log(2\log(\mu/\Lambda_{\MS}))}{4\log^2(\mu/\Lambda_{\MS})} \,+\\
\, & \,+ \frac{\beta_1^2 \log^2(2\log(\mu/\Lambda_{\MS})) - \beta_1^2 \log(2\log(\mu/\Lambda_{\MS}))  +\beta_2\beta_0 - \beta_1^2}{8 \beta_0^5 \log^3(\mu/\Lambda_{\MS})};
\end{aligned}
\eeq
where
\beq
\begin{aligned}
\beta_0 &= \frac{11 }{3}, \qquad &\beta_1 &= \frac{34 }{3}, \qquad &\beta_2 &= \frac{2857 }{54},\\
\gamma_0 &= -\frac{3}{4}, \qquad &\gamma_1 &= -\frac{203}{12}, \qquad
&\gamma_2 &= - \frac{11413}{108}. &
\end{aligned}
\eeq
For the large-$N$ limit of the dynamically-generated scale $\Lambda_{\MS}$, we used the determination of Ref.~\cite{Gonzalez-Arroyo:2012euf} from the TEK model (cf.~also the compatible determination of~\cite{Allton:2008ty} obtained extrapolating from $N\le 10$):
\beq
\frac{\Lambda_{\MS}}{\sqrt{\sigma}} = 0.513(6).
\eeq

On the lattice, apart from the simple bare 't Hooft coupling $\lambda=1/b$ appearing in the TEK Wilson action, one can define improved couplings which are expected to accelerate the convergence of the perturbative series. In this work, we compute the renormalization constant $Z_\P$ both using the simple inverse coupling $b$ appearing in the TEK lattice action, as well as these 3 different definitions of the improved inverse coupling:
\beq\label{eq:improved_couplings_1}
\lambda_{\mathrm{I}} &=& \lambda/P,\\\nonumber
\\[-1em]
\lambda_{\mathrm{E}} &=& 8(1-P),\\\nonumber
\\[-1em]
\lambda_{\mathrm{E}'} &=& -8\log P,
\eeq
where the mean plaquette $P$ is
\beq
P \equiv \frac{1}{6N} \sum_{\mu>\nu} z_{\nu\mu}\left\langle\Tr\left\{U_\mu U_\nu U_\mu^\dagger U_\nu^\dagger\right\}\right\rangle.
\eeq
The improved couplings in Eq.~\eqref{eq:improved_couplings_1} are all equal to $\lambda$ at leading order in perturbation theory, as it can be easily verified using the following perturbative expansion for the mean plaquette~\cite{Athenodorou:2007hi,Bali:2013kia,Perez:2017jyq}:
\beq
P &=& 1 - c_1 \lambda - c_2 \lambda^2 + O(\lambda^3),\\\nonumber
\\[-1em]
c_1 &=& \frac{1}{8},\\\nonumber
\\[-1em]
c_2 &=& 0.0051069297.
\eeq
Since $Z_\P$ is known to two-loop order, in order to compute it using improved couplings we also need the following two-loop relations:
\beq
\lambda&=& \lambda_{\mathrm{I}}-c_1 \lambda_{\mathrm{I}}^2 + \cdots ,\\\nonumber
\\[-1em]
\lambda &=& \lambda_{\mathrm{E}}-\frac{c_2}{c_1}\lambda_{\mathrm{E}}^2 + \cdots,\\\nonumber
\\[-1em]
\lambda &=& \lambda_{\mathrm{E}'} - c_1\left(\frac{1}{2}+\frac{c_2}{c_1^2}\right)\lambda_{\mathrm{E}'}^2 + \cdots .
\eeq

\begin{table}[!t]
\begin{center}
\begin{tabular}{ |c|c|c|c|c|}
\hline
$b$ & $P$ & $1/\lambda_{\mathrm{I}}$& $1/\lambda_{\mathrm{E}}$ & $1/\lambda_{\mathrm{E}'}$ \\
\hline
0.355 & 0.54538(6) & 0.1936 &  0.2750 &  0.2062  \\
0.360 & 0.55800(5) & 0.2009 &  0.2828 &  0.2143  \\
0.365 & 0.56910(5) & 0.2077 &  0.2901 &  0.2217  \\
0.370 & 0.57896(5) & 0.2142 &  0.2969 &  0.2287  \\
\hline
\end{tabular}
\end{center}
\caption{Summary of the definitions of the inverse bare 't Hooft coupling employed to compute $Z_\P$ in the large-$N$ limit from perturbation theory.}
\label{tab:couplings}
\end{table}

\begin{table}[!t]
\begin{center}
\begin{tabular}{|c|c|c|c|c|c|}
\hline
$b$ & \makecell{$Z_\P$\\$[$Pert., $b$$]$} & \makecell{$Z_\P$\\$[$Pert., $bP$$]$} & \makecell{$Z_\P$\\$[$Pert., $\frac{1}{8}(1-P)^{-1}$$]$} & \makecell{$Z_\P$\\$[$Pert., $(-8\log P)^{-1}$$]$} & \makecell{$Z_\P$\\$[$Non-pert.$]$} \\
\hline
0.355 & 0.83423(98) & 0.77944(92) & 0.79278(98) & 0.76896(90) & 0.5212(60) \\
0.360 & 0.82588(81) & 0.77656(76) & 0.78911(82) & 0.76792(76) & 0.536(12)  \\
0.365 & 0.81952(59) & 0.77475(58) & 0.78648(56) & 0.76734(55) & 0.555(12)  \\
0.370 & 0.81487(67) & 0.77376(63) & 0.78476(64) & 0.76732(64) & 0.583(20)  \\
\hline
\end{tabular}
\end{center}
\caption{Determinations of the renormalization constant $Z_\P$ in the large-$N$ limit for the $\MS$ scheme at a renormalization point $\mutwoGeV$ using perturbation theory with various definitions of the inverse bare 't Hooft coupling, and using the non-perturbative determinations of $Z_\S$ of~\cite{Castagnini:2015ejr} plus our non-perturbative determinations for $Z_\P/Z_\S$~\cite{GarciaPerez:2020gnf}.}
\label{tab:ZP}
\end{table}

\begin{figure}[!t]
\centering
\includegraphics[scale=0.42]{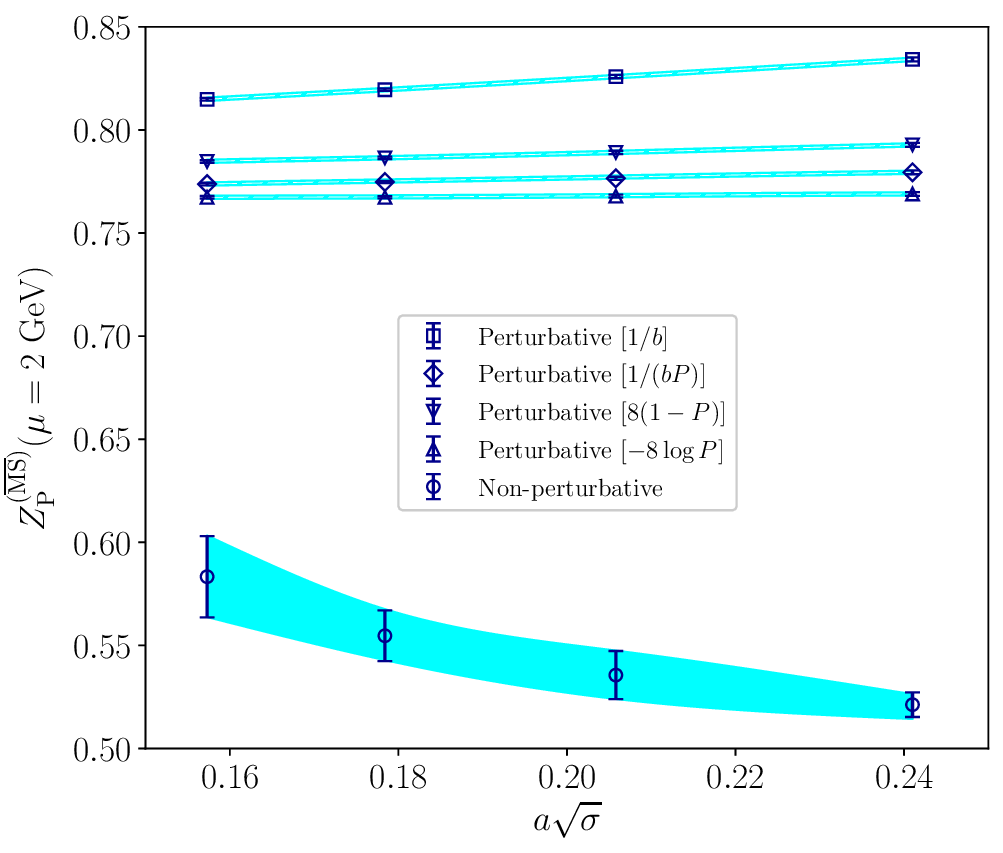}
\caption{Comparison of perturbative large-$N$ calculations of $Z_\P$ using several definitions of the bare 't Hooft coupling, compared to the non-perturbative determinations obtained from results of Ref.~\cite{Castagnini:2015ejr}.}
\label{fig:ZP_perturbative_COMP}
\end{figure}

Now we are able to use Eq.~\eqref{eq:ZP_pert_2loop} to obtain $Z_\P$ from improved couplings. In particular, to do so we just need to replace $\lambda$ by the corresponding expansion, retaining terms up to second order in the improved coupling in the resulting expansion for $Z_P$. In Tab.~\ref{tab:couplings} we report our determinations for the improved inverse couplings for $N=289$, while in Tab.~\ref{tab:ZP} we summarize all determinations of $Z_\P$ discussed so far. We also compare them in Fig.~\ref{fig:ZP_perturbative_COMP}. As it can be appreciated, the naive perturbative calculation largely overestimates the non-perturbative one. Using improved couplings lowers the perturbative estimate, making it closer to the non-perturbative one, and thus going in the expected direction; nonetheless, we still observe a significant overshooting, which is, at best, of about a factor of $\sim 1.3$.

\FloatBarrier

\section{Additional plots of the mode number}\label{app:additional_plots}

\begin{figure}[!htb]
\centering
\includegraphics[scale=0.5]{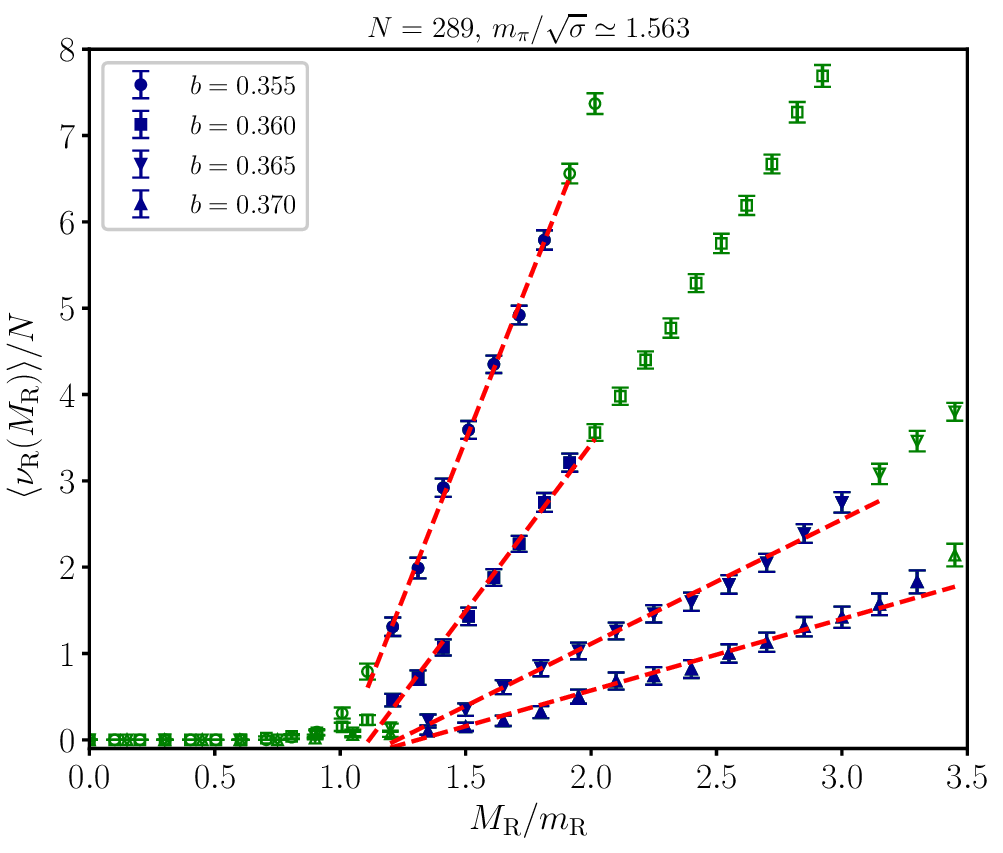}
\includegraphics[scale=0.5]{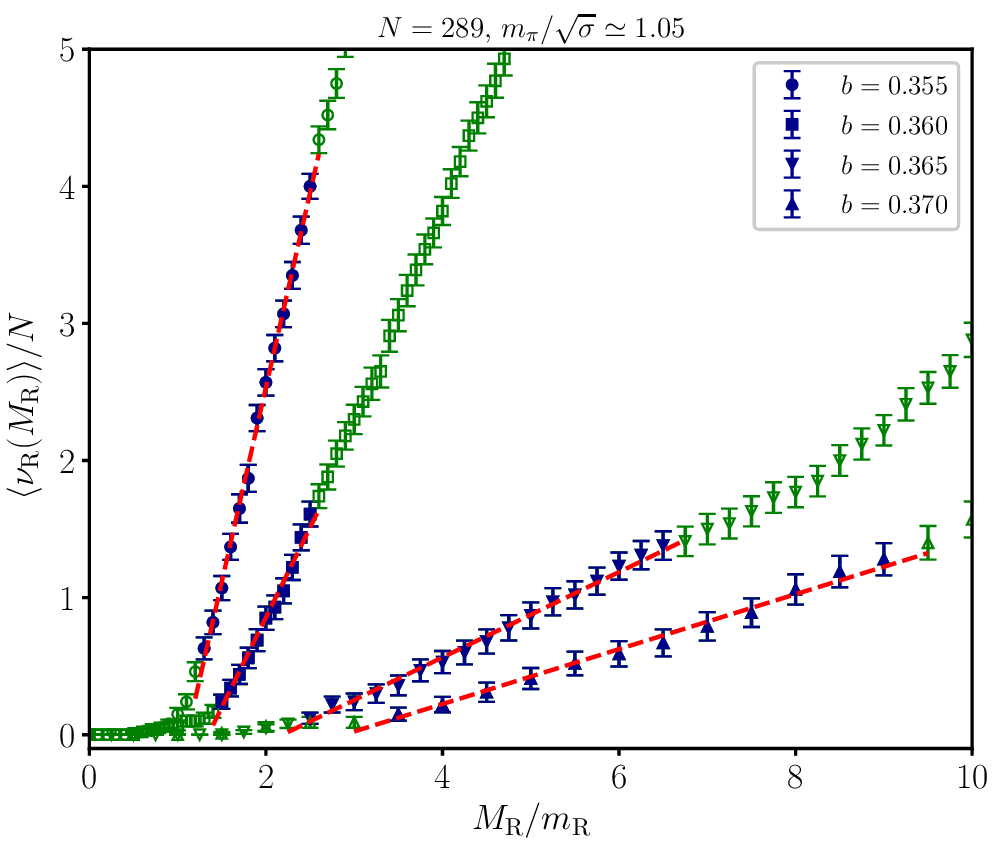}
\caption{Behavior of the mode number $\braket{\nu_\R}/N$, computed for $N=289$, as a function of $M_\R/m_\R$ for four values of the bare coupling $b$ and four different values of $\kappa$, approximately corresponding to $m_\pi/\sqrt{\sigma} \simeq 1.56$ (top) and $m_\pi/\sqrt{\sigma} \simeq 1.05$ (bottom). Full points correspond to those included in the linear best fit to determine the slope.}
\label{fig:additional_plots_mode_number}
\end{figure}

In Fig.~\ref{fig:additional_plots_mode_number} we show two additional plots about the linear best fit of the mode number for $m_\pi/\sqrt{\sigma} \simeq 1.536$ and $1.050$, not shown in the main text.

\FloatBarrier

\providecommand{\href}[2]{#2}\begingroup\raggedright\endgroup

\end{document}